\documentclass[prd,twocolumn,showpacs,nofootinbib]{revtex4}
\usepackage{times}
\usepackage{natbib}
\usepackage{epsfig}

\newcommand{\ang}{\hat\nabla}

\newcommand{\bdv}[1]{{\bf #1}}

\newcommand{\up}[1]{{\rm #1}}
\newcommand{\kms}{{\rm km\, s}^{-1}}
\newcommand{\mpc}{{\rm Mpc}}
\newcommand{\hmpc}{{h^{-1}\mpc}}
\newcommand{\kpc}{{\rm kpc}}
\newcommand{\hkpc}{{h^{-1}\kpc}}
\newcommand{\msun}{M_{\odot}}
\newcommand{\hmsun}{{h^{-1}\msun}}
\newcommand{\Rvir}{R_\up{vir}}
\newcommand{\OM}{\Omega_m}
\newcommand{\OB}{\Omega_b}
\newcommand{\OL}{\Omega_\Lambda}
\newcommand{\rms}{\sigma_8}

\newcommand{\beeq}{\begin{equation}} 
\newcommand{\eneq}{\end{equation}}
\newcommand{\bear}{\begin{eqnarray}}
\newcommand{\enar}{\end{eqnarray}}

\newcommand{\vL}{\bdv{l}}
\newcommand{\Vang}{\bdv{\hat n}}
\newcommand{\Sang}{\bdv{\hat s}}
\newcommand{\lenT}{\tilde T}
\newcommand{\lenQ}{\tilde Q}
\newcommand{\lenU}{\tilde U}
\newcommand{\lenE}{\tilde E}
\newcommand{\lenB}{\tilde B}
\newcommand{\CT}{C^T}
\newcommand{\CE}{C^E}
\newcommand{\CB}{C^B}
\newcommand{\CC}{C^C}
\newcommand{\lenCT}{\tilde C^T}
\newcommand{\lenCE}{\tilde C^E}
\newcommand{\lenCB}{\tilde C^B}
\newcommand{\lenCC}{\tilde C^C}
\newcommand{\Cphi}{C^\phi}
\newcommand{\Tobs}{\tilde T^\up{obs}}
\newcommand{\CTobs}{\tilde C^{T,\up{obs}}}
\newcommand{\Xobs}{\tilde X^\up{obs}}
\newcommand{\CXobs}{\tilde C^{X,\up{obs}}}
\newcommand{\Yobs}{\tilde Y^\up{obs}}
\newcommand{\CYobs}{\tilde C^{Y,\up{obs}}}
\newcommand{\GXY}{\bdv{G}_{XY}}
\newcommand{\WY}{W_Y}
\newcommand{\khat}{\hat\kappa}
\newcommand{\Ckap}{C^\kappa}
\newcommand{\Nkap}{N^{\kappa,XY}}
\newcommand{\Kmod}{\kappa^m}
\newcommand{\Pmod}{\phi^m}
\newcommand{\Tksz}{T_\up{kSZ}}
\newcommand{\muK}{\mu\up{K}}
\newcommand{\Scrit}{\Sigma_\up{crit}}
\newcommand{\vlos}{v_\up{los}}
\newcommand{\YHe}{Y_\up{He}}
\newcommand{\spix}{\sigma_\up{pix}}
\newcommand{\opix}{\Omega_\up{pix}}

\newcommand{\sbeam}{\sigma_b}
\newcommand{\tfwhm}{\theta_\up{FWHM}}

\begin{document}

\title{Lensing reconstruction of cluster-mass cross-correlation
with cosmic microwave background polarization}

\author{Jaiyul Yoo$^1$}
\altaffiliation{jyoo@cfa.harvard.edu} 
\author{Matias Zaldarriaga$^{2,1,3}$}
\author{Lars Hernquist$^1$}
\affiliation{$^1$Harvard-Smithsonian Center for Astrophysics, Harvard 
University, 60 Garden Street, Cambridge, MA 02138}
\affiliation{$^2$School of Natural Sciences,
Institute for Advanced Study, Einstein Drive, Princeton, NJ 08540}
\affiliation{$^3$Jefferson Physical Laboratory, Harvard University, 
17 Oxford Street, Cambridge, MA 02138}

\begin{abstract}
We extend our maximum likelihood method for reconstructing the cluster-mass 
cross-correlation from cosmic microwave background (CMB) temperature 
anisotropies and develop new estimators that utilize
six different quadratic combinations of CMB temperature and polarization
fields. Our maximum likelihood estimators are constructed with delensed
CMB temperature and polarization fields by using an assumed model of the
convergence field and they can be iteratively applied to a set of clusters,
approaching to the optimal condition for the lensing reconstruction
as the assumed initial model is refined.
Using smoothed particle hydrodynamics simulations, we create a catalog of
realistic clusters obtainable from the current Sunyaev-Zel'dovich (SZ) surveys,
and we demonstrate the ability of the maximum likelihood estimators to
reconstruct the cluster-mass cross-correlation from the massive clusters.
The iTT temperature estimator provides a signal-to-noise ratio of a factor 3
larger than the iEB polarization estimator, unless the detector noise for
measuring polarization anisotropies is controlled under $3\muK$. 
\end{abstract}


\maketitle

\section{Introduction}
In the past few years we have seen rapid development in the 
measurements of cosmic
microwave background (CMB) anisotropies on arcminute scales. 
Higher precision measurements of CMB temperature anisotropies are available
from the completed missions such as the Arcminute Cosmology Bolometer Array 
Receiver \cite{READET09} and the Cosmic Background Imager \cite{SIMAET09}
and the ongoing experiments with better angular resolution and multi-frequency
channels
such as the Atacama Cosmology Telescope (ACT, \cite{HIACET09}) and
the South Pole Telescope (SPT, \cite{STADET08}). While most of the current
polarization experiments are aimed at measuring CMB polarization 
anisotropies on large scales, ACT and SPT will be capable of measuring
CMB polarization anisotropies on arcminute scales.

Especially, ACT and SPT are designed for measuring the Sunyaev-Zel'dovich (SZ)
effect arising from the Thomson scattering off hot electrons in clusters
and thereby detecting massive clusters.
These SZ surveys can explore the growth of structure by measuring the 
abundance of massive clusters and the expansion history of the universe by
measuring their correlation functions to probe the evolution of dark energy
(see, e.g., \cite{DETF06}). 
With precisely measured distance and their omnipresence behind all the
clusters, CMB anisotropies can be used as a distant background source 
for weak lensing measurements, providing the cluster-mass cross-correlation
and complementing the measurements of cluster mass and their abundance from
the same SZ surveys. Moreover, weak lensing measurements of the CMB can be
combined with galaxy weak lensing measurements of the same clusters from
optical follow-up surveys to measure the source distance ratios in a 
model-independent way, constraining the expansion history of the universe 
\cite{HUHOVA07}.

Weak lensing of the CMB by large-scale structure is efficiently probed by using
quadratic estimators, measuring the deviation of the correlation function
of CMB anisotropies from the otherwise statistically isotropic correlation
\cite{HU01b}. However, it was shown that
the standard method of using quadratic estimators
is compromised \cite{MABAMEET05} on cluster scales
and needs additional free parameters \cite{HUDEVA07} for calibration, as 
this method rests on the linear approximation in the lensing effect which
breaks down near massive clusters.
This problem was tackled by delensing the observed CMB temperature field
and analyzing the likelihood of the delensed fields \cite{YOZA08}.

We showed that if the assumed initial model for delensing is a good 
approximation to the true underlying matter distribution, our new estimator
based on the delensed CMB temperature fields becomes an optimal estimator
and it can be iteratively applied to given measurements of CMB temperature
anisotropies, until the assumed initial model converges to the true matter
distribution and the likelihood is maximized.
Here we extend our maximum likelihood estimation method for reconstructing the
cluster-mass cross-correlation with CMB temperature anisotropies and 
apply it to CMB polarization anisotropies, yielding new maximum likelihood
estimators with six different quadratic combinations of CMB temperature and
polarization fields. We demonstrate their applicability 
by using realistic clusters that can be found in the current CMB experiments.

The rest of the paper is organized as follows. We briefly review the formalism
for weak lensing of CMB anisotropies and derive our maximum likelihood
estimators in Sec.~\ref{sec:for}. In Sec.~\ref{sec:sim} we describe our
numerical simulations to model realistic clusters and we 
construct a catalog of massive clusters with CMB secondary anisotropies.
We reconstruct the cluster-mass cross-correlation
by using our maximum likelihood estimators and quantify their signal-to-noise
ratios in Sec.~\ref{ssec:iqe}, and we investigate the impact of the kSZ
contamination in Sec.~\ref{ssec:ksz}. We summarize our findings and
conclude in Sec.~\ref{sec:dis}. For illustrative purposes,
we adopt a flat $\Lambda$CDM universe with cosmological parameters
($\OM h^2=0.127$, $\OB h^2=0.0222$, $h=0.73$, $n_s=0.95$, $\rms=0.78$),
consistent with the recent estimation (e.g., \citep{TESTET06,KODUET08}).

\section{Formalism}
\label{sec:for}
\subsection{Weak lensing of CMB polarization}
\label{ssec:weak}
Gravitational lensing is a surface brightness conserving process and it
simply redistributes the intrinsic CMB temperature and polarization fields.
Here we use $T$, $Q$, and $U$ to represent the CMB temperature and the Stokes
parameters for the
CMB polarization in units of $\muK$ and use tildes to indicate
that the corresponding field is gravitationally lensed. The lensed CMB 
temperature and polarization fields
at the angular position $\Vang$ on the sky are therefore
\bear
\label{eq:lensing}
\lenT(\Vang)&=&T\left[\Vang+\ang\phi(\Vang)\right]~,\\
(\lenQ\pm i\lenU)(\Vang)&=&(Q\pm iU)\left[\Vang+\ang\phi(\Vang)\right]~,
\nonumber
\enar
and the projected potential 
\beeq
\phi(\Vang)=-2\int_0^{D_\star}dD~{D_\star-D\over D D_\star}~\psi~(D\Vang,D)~
\eneq
describes the deflection angle $\ang\phi(\Vang)$ on the sky,
where the gravitational
potential is $\psi$, the comoving angular diameter distance to the CMB
last scattering surface is $D_\star=14.12$~Gpc \cite{KODUET08},
and $\ang$ is the derivative with respect to unit angular vector $\Vang$.
The projected potential~$\phi$
is related to the convergence $\kappa$ 
as $\ang^2\phi=-2\kappa$ and it is further related to the physical matter
density~$\delta\rho_m$ projected along the line-of-sight as
\beeq
\kappa(\Vang)=\int_0^{D_\star}dD~{D(D_\star-D)\over (1+z)^2D_\star}
~4\pi G~\delta\rho_m(D\Vang,D)~.
\eneq

As we are interested in reconstructing the cluster-mass cross-correlation,
 we adopt the
flat sky approximation and express quantities of interest in Fourier space.
To linear order in~$\phi$, the lensed CMB temperature field can be 
written as
\beeq
\label{eq:lenT}
\lenT_\vL=T_\vL-\int{d^2\vL_1\over(2\pi)^2}~(\vL_2\cdot\vL_1)
~\phi_{\vL_2}~T_{\vL_1}~,
\eneq
with $\vL_2=\vL-\vL_1$
and this equation is also valid for the lensed Stokes parameters
$\lenQ_\vL$~and~$\lenU_\vL$ in Fourier space.
However, since density fluctuations excite only curl-free polarization,
the CMB polarization fields are better described
by two parity eigenstates, $E$-~and~$B$-modes \cite{SEZA97} as
$(Q_\vL\pm iU_\vL)=(E_\vL\pm iB_\vL)e^{\pm2i\varphi_\vL}$ with the phase
$\varphi_\vL$ of the wavevector~$\vL$.
The lensed CMB polarization fields are
therefore
\bear
\label{eq:lenEB}
\lenE_\vL&=&E_\vL-\int{d^2\vL_1\over(2\pi)^2}~(\vL_2\cdot\vL_1)
~\phi_{\vL_2}~E_{\vL_1}\cos2\Delta\varphi_{\vL_1}~,  \\
\lenB_\vL&=&-\int{d^2\vL_1\over(2\pi)^2}~(\vL_2\cdot\vL_1)
~\phi_{\vL_2}~E_{\vL_1}\sin2\Delta\varphi_{\vL_1}~, \nonumber
\enar
and the lensed CMB power spectra are related to the intrinsic CMB power
spectra as
\bear
\lenCT_l&=&\left[1-l^2R\right]\CT_l+\int{d^2\vL_1\over(2\pi)^2}
~(\vL_2\cdot\vL_1)^2~\Cphi_{l_2}~\CT_{l_1}~, \\
\lenCE_l&=&\left[1-l^2R\right]\CE_l+\int{d^2\vL_1\over(2\pi)^2}~
(\vL_2\cdot\vL_1)^2~\Cphi_{l_2}~\CE_{l_1}\cos^22\Delta\varphi_{\vL_1},\nonumber \\
\lenCC_l&=&\left[1-l^2R\right]\CC_l+\int{d^2\vL_1\over(2\pi)^2}~
~(\vL_2\cdot\vL_1)^2~\Cphi_{l_2}~\CC_{l_1}\cos^22\Delta\varphi_{\vL_1},\nonumber\\
\lenCB_l&=&\int{d^2\vL_1\over(2\pi)^2}
~(\vL_2\cdot\vL_1)^2~\Cphi_{l_2}~\CE_{l_1}\sin^22\Delta\varphi_{\vL_1}~, \nonumber
\enar
where $\Delta\varphi_{\vL_1}=\varphi_{\vL_1}-\varphi_\vL$ and
$R=(1/4\pi)\int d\ln l~l^4\Cphi_l$ 
is the half of the rms deflection angle (see, e.g., \cite{HU00,LECH06}).

Finally, we assume the Gaussian random noise for the detector and the Gaussian
beam of the telescope. Therefore, the power spectrum of the detector noise
is \cite{KNOX95}
\beeq
C_l^{N,T}=\spix^2~\opix~,
\label{eq:nois}
\eneq
and the observed CMB temperature field and its power spectrum are
\bear
\label{eq:tobs}
\Tobs_\vL&=&\lenT_\vL~e^{-{1\over2}l^2\sbeam^2}+N^T_\vL~,\\
\CTobs_l&=&\lenCT_le^{-l^2\sbeam^2}+C^{N,T}_l~, \nonumber
\enar
where 
$\spix$ is the rms noise of the detector, $\opix$ is the solid angle
subtending each pixel of the detector, $N^T_\vL$ is the Fourier mode of the
detector noise, and
the full-width half-maximum (FWHM) of the telescope beam is
$\tfwhm=\sbeam\sqrt{8\ln2}$~. The observed CMB polarization fields are 
also described
by Eqs.~(\ref{eq:tobs}), while $\spix$ and $\sbeam$ in CMB polarization
experiments may differ from those in CMB experiments for measuring 
temperature anisotropies.

\subsection{Improved quadratic estimators}
\label{ssec:iqes}
Matter fluctuations along the line-of-sight deflect CMB photons and
this process imprints a deviation of the CMB two-point statistics from
statistical isotropy. Quadratic estimators are often used to measure
the deviation, but they become progressively biased as the lensing effect
increases. This problem can be overcome by using improved quadratic estimators
that are constructed by using delensed CMB fields \cite{YOZA08}.
Here we briefly review the standard quadratic estimators \cite{HUDEVA07}
using the CMB  polarization fields, on which our
improved quadratic estimators are based, and then we describe how our 
improved quadratic estimators can be used to reconstruct the cluster-mass
cross-correlation.

With the CMB temperature and polarization fields, we can construct six
convergence estimators $\khat^{XY}(\Vang)$ that use two CMB temperature and
polarization fields $X,Y=T,E,B$, permitting repetition (hence they are 
quadratic estimators). 
The estimators $\khat^{XY}(\Vang)$ are symmetric with two fields
$X$~and~$Y$ interchanged and we use quantities with a hat to indicate that
they are estimators thereof (not to be confused with unit angular vectors
on the sky).
The six convergence estimators are uniquely determined with the requirement
that each estimator be unbiased 
$\langle\khat^{XY}(\Vang)\rangle=\kappa(\Vang)$~
over an ensemble average of the CMB temperature and polarization
fields $X$~and~$Y$, 
and the variance of the estimator be minimal \cite{HU01b},
\beeq
\langle\khat^{XY}_\vL\khat^{*XY}_{\vL'}\rangle=(2\pi)^2~\delta^D(\bdv{l-l'})
(\Ckap_l+\Nkap_l)~.
\label{eq:vari}
\eneq
These estimators can be easily computed in configuration
space by using the two  Wiener-filtered functions
\bear
\label{eq:filter}
&&\GXY(\Vang)=\int\!\!\!{d^2\vL\over(2\pi)^2}~i\vL\Xobs_\vL
{C_l^{XY}\over\CXobs_l}
\bigg\{\begin{array}{c}e^{2i\varphi_{\vL}}\\e^{2i\varphi_{\vL}}\end{array}\bigg\}
e^{-{1\over2}l^2\sbeam^2+i\vL\cdot\Vang}~, \nonumber \\
&&\WY(\Vang)=\int\!\!\!{d^2\vL\over(2\pi)^2}~{\Yobs_\vL\over\CYobs_l}
\bigg\{\begin{array}{c}e^{2i\varphi_{\vL}}\\ie^{2i\varphi_{\vL}}\end{array}\bigg\}
e^{-{1\over2}l^2\sbeam^2+i\vL\cdot\Vang}~,
\enar
with the phase angle  $\varphi_\vL$ of the wavevector~$\vL$ and
two phase factors for $Y=E,B$ in the braces. Note that $C_l^{XY}=C_l^{XE}$
for $Y=B$ and the phase factor in the braces is unity for $Y=T$.
The convergence estimator is
then
\beeq
\khat^{XY}_\vL=-{N^{XY}_l\over2}~i\vL\cdot\int d^2\Vang~
\up{Re}\left[\GXY(\Vang)\WY^*(\Vang)\right]
~e^{-i\vL\cdot\Vang}~.
\label{eq:GW}
\eneq
The normalization coefficients $N^{XY}_l$ are related to the noise power
spectrum $\Nkap_l$ of XY estimators $\khat^{XY}(\Vang)$ as
$\Nkap_l=l^2N^{XY}_l/4$, and they can be obtained as
\bear
\label{eq:nps}
{1\over N^{XY}_l}&=&{1\over l^2}\int{d^2\vL_1\over(2\pi)^2}{(\vL\cdot\vL_1)~
C^{XY}_{l_1}f^{XY}_{\vL_1\vL_2}\over\CXobs_{l_1}~\CYobs_{l_2}} \\
&\times&
\bigg\{\begin{array}{c}\cos2\Delta\varphi\\\sin2\Delta\varphi\end{array}\bigg\}
~e^{-l_1^2\sbeam^2}~e^{-l_2^2\sbeam^2}~,\nonumber
\enar
with $\vL=\vL_1+\vL_2$, $\Delta\varphi=\varphi_{\vL_1}-\varphi_{\vL_2}$,
and $\langle X_{\vL_1}Y_{\vL_2}\rangle=f_{\vL_1\vL_2}^{XY}~\phi_\vL$~,
where
\bear
f^{TT}_{\vL_1,\vL_2}&=&(\vL\cdot\vL_1)~\CT_{l_1}+(\vL\cdot\vL_2)~\CT_{l_2}~, \\
f^{TE}_{\vL_1,\vL_2}&=&(\vL\cdot\vL_1)~\CC_{l_1}\cos2\Delta\varphi
+(\vL\cdot\vL_2)~\CC_{l_2}~, \nonumber \\
f^{TB}_{\vL_1,\vL_2}&=&(\vL\cdot\vL_1)~\CC_{l_1}\sin 2\Delta\varphi~,\nonumber \\
f^{EE}_{\vL_1,\vL_2}&=&\left[(\vL\cdot\vL_1)~\CE_{l_1}+(\vL\cdot\vL_2)~\CE_{l_2}
\right]\cos2\Delta\varphi~,\nonumber \\
f^{EB}_{\vL_1,\vL_2}&=&(\vL\cdot\vL_1)~\CE_{l_1}\sin2\Delta\varphi~\nonumber
\enar
(see \cite{HUOK02,HUDEVA07} for details). Due to the vanishing signal-to-noise
ratio, no quadratic BB estimator is used.

Improved quadratic estimators are similar in many aspects to the above
standard quadratic estimators. Our estimation process is as follows:
Improved quadratic estimators take an initial model $\Kmod(\Vang)$
of the convergence field and we first
compute the delensed CMB temperature and polarization
fields $X^d(\Sang)=\Xobs(\Vang)$ by solving the
lensing equation $\Sang=\Vang+\ang\Pmod(\Vang)$, where the initial model
$\phi^m(\Vang)$ for the projected potential is related to $\Kmod(\Vang)$
as $\ang^2\Pmod(\Vang)=-2~\Kmod(\Vang)$~. Improved quadratic estimators
are constructed by using the same Wiener-filtered functions
$\GXY(\Vang)$ and $\WY(\Vang)$ but with the delensed CMB temperature
and polarization fields $X^d$ and $Y^d$ rather than
$\Xobs$ and $\Yobs$ themselves, and Eq.~(\ref{eq:GW}) yields the change
$\Delta\kappa^{XY}(\Vang)$ in the convergence field
with respect to the assumed initial model
$\Kmod(\Vang)$. The resulting convergence field
$\Kmod(\Vang)+\Delta\kappa^{XY}(\Vang)$ then serves as a refined 
initial model for the next iteration and this process is iterated until
the numerical convergence is achieved 
$\Delta\kappa^{XY}(\Vang)/\Kmod(\Vang)\ll1$. 
In practice, one can try different initial models for a faster numerical
convergence, when $\Delta\kappa^{XY}(\Vang)\simeq\Kmod(\Vang)$.

While our improved quadratic estimators take the same functional form
as the standard quadratic estimators (thereby its name is inherited),
there exist critical differences:
Our improved quadratic estimators are based on the maximum likelihood of
the delensed CMB temperature and polarization fields, and the iteration process
is indeed the standard Newton-Raphson method for maximizing the likelihood
of each estimate given the observed
CMB temperature and polarization fields. Since
the initial model $\Kmod(\Vang)$
that depends on the previous iteration process is refined and 
the refined model $\Kmod+\Delta\kappa$ becomes a new
initial model in the following iteration,
the improved quadratic estimators are in fact 
{\it rational} functions of the CMB temperature and 
polarization fields, 
rather than quadratic functions, making full use of the information 
contained in the likelihood.

Furthermore, our improved quadratic estimators are free from the 
approximation that the lensing effect is weak,
the breakdown of which plagues the standard quadratic estimators. 
Consequently, 
there is no arbitrary cutoff scale used in modified quadratic
estimators \cite{HUDEVA07}
that breaks the symmetry in the convergence estimators when
the CMB temperature and polarization fields $X$~and~$Y$ are interchanged.
Hereafter, we refer to improved quadratic estimators using delensed CMB 
temperature and polarization fields $X$~and~$Y$ as iXY estimators.

\section{Massive Clusters}
\label{sec:sim}
We describe our model for massive clusters in Sec.~\ref{ssec:num} and 
discuss possible contaminants for lensing reconstruction 
arising from secondary anisotropies in the CMB temperature 
and polarization fields in Sec.~\ref{ssec:sec}.
Tests of the applicability of our improved quadratic estimators to realistic
clusters are presented in Sec.~\ref{sec:recon}.

\subsection{Numerical simulation}
\label{ssec:num}
Here we model massive clusters using the numerical simulations of
\citet{SPHE03}.
We use the largest volume simulation (G-series) among a number of
simulation runs with varying mass and spatial resolution.
The smoothed particle hydrodynamics (SPH) simulations were performed
using the parallel GADGET code \cite{GADGET}, employing the 
variational formulation of \citet{SPHE02}. 
We focus on the simulation
with $2\times324^3$ of gas and dark matter particles in a comoving cubic volume
$(100\hmpc)^3$. 
Dark matter halos are identified by applying the friends-of-friends (FoF) group
finding algorithm \cite{DEFW85} to the dark matter distribution with a 
comoving linking length of 0.2 times the mean interparticle separation
$62\hkpc$. Although the simulation was run in a flat $\Lambda$CDM universe
with slightly different cosmological parameters ($\OM=0.3$, 
$\OL=0.7$, $\OB=0.04$, $n_s=1$, $\rms=0.9$), we simply
retain the physical properties
of the simulation such as particle mass, position, and velocity.

\begin{figure}
\centerline{\psfig{file=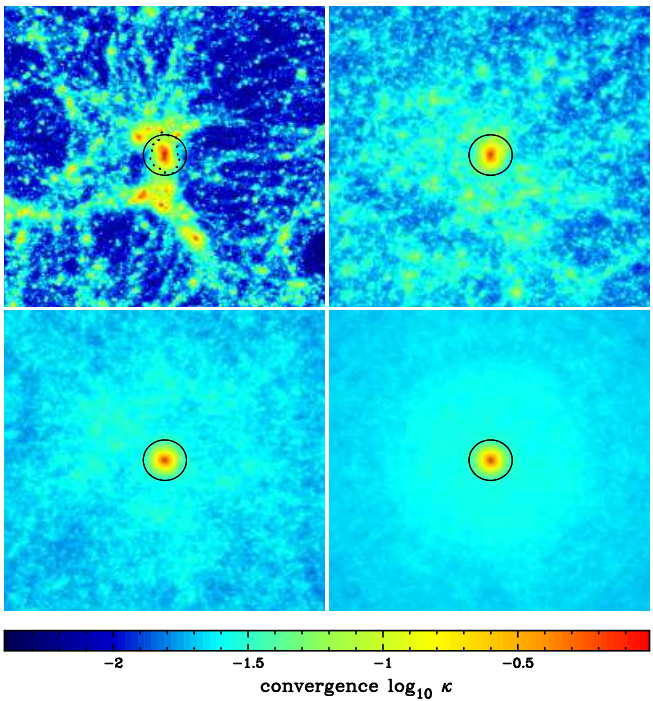, width=3.4in}}
\caption{(color online) Average convergence fields $\kappa(\Vang)$ of a
$40'\times40'$ region around massive clusters at $z=1$. Each panel shows
$\kappa(\Vang)$ obtained by averaging 1 (top left), 10 (top right), 
100 (bottom left), and 1000 (bottom right) identical but
randomly rotated regions around the massive cluster from the smoothed
particle hydrodynamics simulation, which represents different clusters
of the same mass. The cluster has of mass 
$M=3.7\times10^{14}\hmsun$ and virial radius $\Rvir=1.9\hmpc$. Circles
show $\Rvir$ on the sky subtended by 2$.\!'$7 and the dotted line in the
upper left panel shows the boundary of the cluster members identified by the
friends-of-friends algorithm. Once many lines-of-sight
are stacked, the average convergence field restores the spherical symmetry.}
\label{fig:stack}
\end{figure}

\begin{figure}
\centerline{\psfig{file=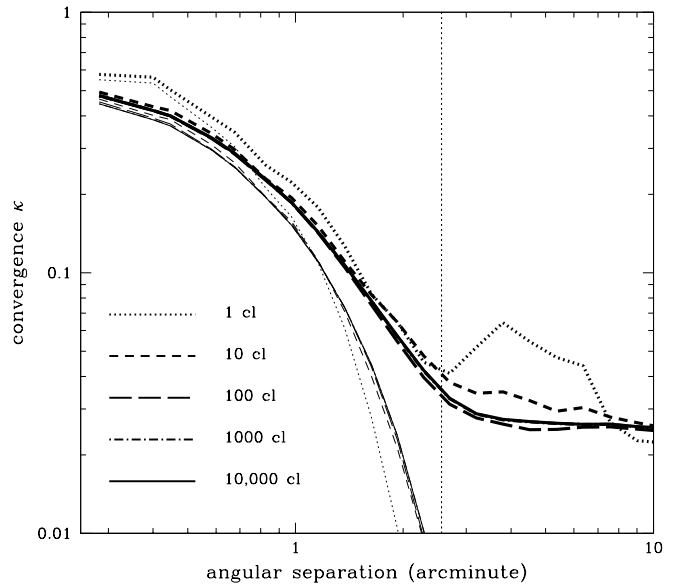, width=3.4in}}
\caption{Convergence profiles $\kappa(\theta)$ around massive clusters at $z=1$.
Profiles are obtained by averaging the stacked convergence fields
$\kappa(\Vang)$ in Fig.~\ref{fig:stack} over the annulus~$\theta$
around the cluster
center. Thin and thick lines show the convergence profiles obtained by 
counting only the cluster members identified by the
friends-of-friends algorithm and by counting all the particles within the
annulus, respectively. The thick solid line
represents the cluster-mass cross-correlation
and it deviates substantially from the average cluster mass profile 
(thin solid) around the virial radius. The average cluster mass profile
is well approximated by a projected NFW profile truncated at the virial
radius (vertical dotted). Note that dot-dashed lines are largely obscured
by solid lines, indicating fast convergence of $\kappa(\theta)$.}
\label{fig:prof}
\end{figure}

To generate a catalog of massive clusters that can be found in the
current Sunyaev-Zel'dovich surveys, we first select the most massive halo
in the simulation output at $z=1$; The massive halo contains $\sim$150,000
dark matter particles, corresponding to virial mass $M=3.7\times10^{14}\hmsun$ 
and radius $\Rvir=1.9\hmpc$. Every particle in the simulation is then
shifted to have the massive halo at the center of the simulation box
by using the periodic
boundary condition, and the whole simulation box
is randomly rotated to provide the distant
observer with different lines-of-sight to the massive cluster.
We treat the massive cluster seen at different lines-of-sight as
independent clusters of the same mass at $z=1$. This process is necessary
to generate a large number of massive clusters using SPH simulations.
Since individual clusters identified by the FoF algorithm lack
spatial symmetry, snapshots of the cluster at different lines-of-sight are
relatively independent of each other as far as the projected matter density
is concerned.

Figure~\ref{fig:stack} shows the convergence fields $\kappa(\Vang)$ of a
$40'\times40'$ region on the sky around the massive cluster with the virial
radius indicated as the circle at the center. 
With the fixed distance $D_\star$ to the background source, the critical
(physical) surface density is just a function of the redshift of the lensing
cluster $1/\Scrit=4\pi GD(D_\star-D)/D_\star(1+z)$
and the convergence is the ratio of the projected mass density $\Sigma(\Vang)$
to the critical surface density $\kappa(\Vang)=\Sigma(\Vang)/\Scrit$~.
The comoving angular diameter distance to the redshift slice $z=1$
is $D=2400\hmpc$ and the corresponding critical surface density is
$\Scrit=1800~h\msun\up{pc}^{-2}$. Here we keep fixed the detector scale
$0.\!'2$ per pixel, smaller than the typical telescope beam size.
Each dark matter particle in the simulation
is of mass $m_\up{dm}=2.1\times10^9\hmsun$ and it contributes
$\kappa=6.0\times10^{-5}$ per pixel$^2$ at $z=1$. 
Many lines-of-sight are randomly generated
and each panel shows $\kappa(\Vang)$ averaged over 1, 10, 100, and 1000
different lines-of-sight to the massive cluster, mimicking the process of
stacking many different clusters at $z=1$.
Once many clusters are stacked, the
average convergence field (bottom right) restores the spherical symmetry.

Figure~\ref{fig:prof} shows the convergence profiles $\kappa(\theta)$ averaged
over annulus as a function of angular separation~$\theta$ from the cluster
center. Each line represents $\kappa(\theta)$ from the average convergence
fields obtained by stacking as many clusters indicated in the legend.
Projected NFW profiles with truncation at $\Rvir$ provide a good approximation
to the average cluster mass profile (thin solid) obtained by counting only
the cluster members identified by the FoF algorithm. However, weak lensing
measures the projected mass distribution including contributions from 
interlopers that happen to lie between the lensing cluster and the observer.
The average convergence profile from the stacked field, therefore, provides
the cluster-mass cross-correlation (thick solid).

\begin{figure}
\centerline{\psfig{file=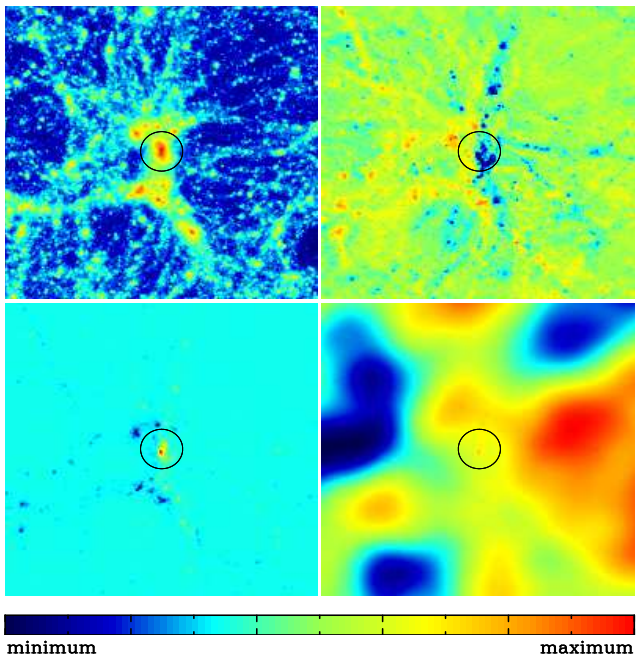, width=3.4in}}
\caption{(color online) Kinetic Sunyaev-Zel'dovich (kSZ) effect.
Top left: Thomson scattering optical depth~$\tau$ of the same region 
in Fig.~\ref{fig:stack}. We assume the helium mass fraction is $\YHe=0.24$ and
helium atoms are doubly ionized.
Top right: Line-of-sight velocity field $\vlos$ of dark matter particles.
Bottom left: The kSZ effect $\Delta\Tksz/T=-\tau~\vlos/c$. 
Electrons are assumed to follow the dark matter distribution. 
Bottom right: $\Delta\Tksz$ from the kSZ effect
is superimposed to the intrinsic CMB temperature field.
The horizontal color bar represents the scales of each panel in the range
of $-4.4\leq\log\tau\leq-2.0$, $-700~\kms\leq \vlos\leq560~\kms$,
$-16~\muK\leq\Delta\Tksz\leq30~\muK$, and 
$-148~\muK\leq\Delta T\leq110~\muK$, respectively.}
\label{fig:ksz}
\end{figure}

\begin{figure*}
\centerline{\psfig{file=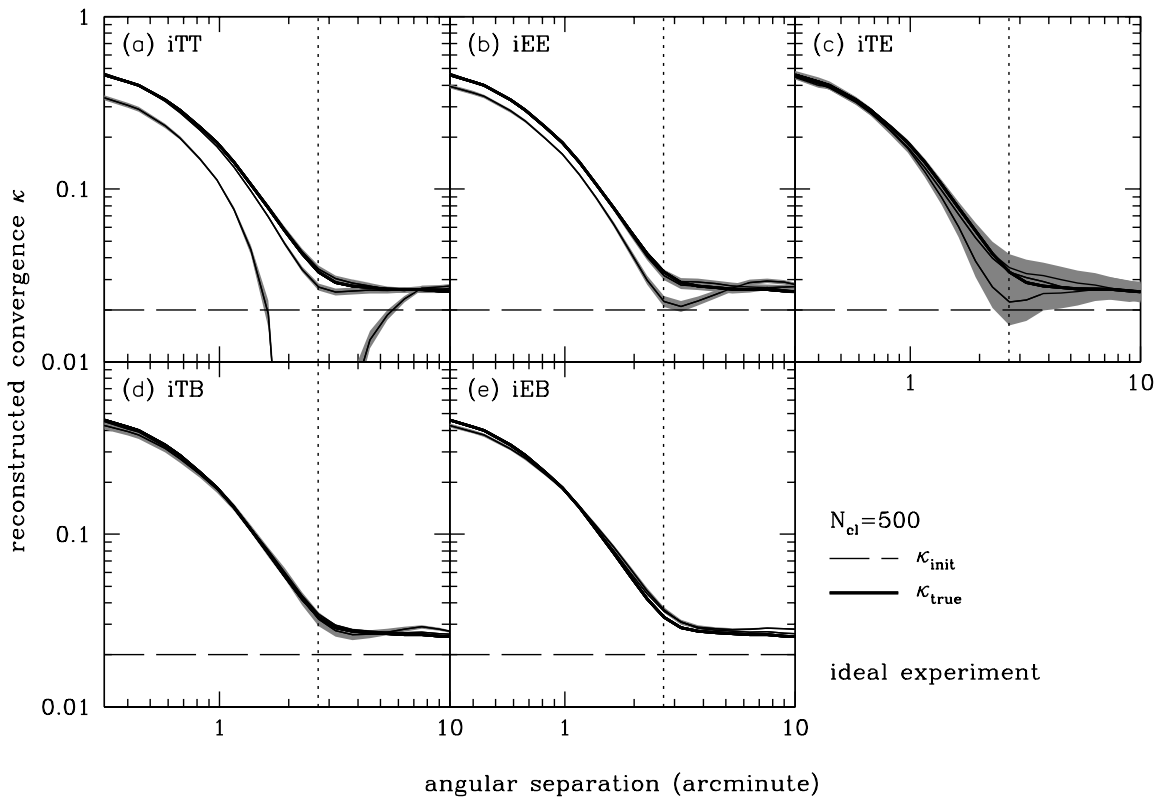, width=6in}}
\caption{Lensing reconstruction of the convergence profile~$\kappa(\theta)$
using improved quadratic 
estimators based on CMB temperature and polarization fields in an ideal CMB
experiment ($\spix=\tfwhm=0$).
The initial model for the convergence field is assumed to have
a uniform matter distribution $\Kmod(\theta)=0.02$ (dashed) and 
improved quadratic estimators are applied to 500 clusters, yielding
 an estimate of the convergence profile (thin solid) with its uncertainty
(shaded) at each iteration, and refining 
the initial input model. All the improved quadratic estimators converge
quickly to the true
convergence profile (thick solid) in a few iterations.}
\label{fig:ideal}
\end{figure*}

\subsection{Secondary anisotropies by clusters}
\label{ssec:sec}
On cluster scales, the CMB temperature and polarization fields can be 
approximated as large-scale gradient fields, and their gradient directions are 
nearly uncorrelated. Gravitational lensing 
by massive clusters induce dipolelike wiggles around the clusters
to the CMB temperature and polarization gradient fields, and
these features are used to measure the lensing effect \cite{SEZA95}. 
However, there exist several sources of contamination that 
mimic the lensing signature and complicate its reconstruction process.

Hot electrons in massive clusters scatter off CMB photons, giving rise
to a spectral distortion of the otherwise Planck distribution of the
CMB temperature field. This thermal Sunyaev-Zel'dovich (tSZ) effect 
\cite{SUZE70a}
can be as large as $\Delta T\simeq500\muK$ and it is a major source of
contamination, significantly exceeding the lensing signals from the clusters.
However, with its distinctive spectral signature the tSZ effect can be
removed by multi-frequency observations and here we ignore the possible 
contamination of the residual tSZ effect arising from
an imperfect cleaning process. However,
the scattering of hot electrons in massive
clusters also gives rise to the Doppler effect due to the bulk motion
of the clusters. Though smaller than the tSZ effect, this kinetic
Sunyaev-Zel'dovich (kSZ) effect \cite{SUZE72} is comparable to the lensing
signals $\Delta\Tksz\simeq30\muK$ and its identical spectral dependence
makes it hard to separate from the intrinsic CMB temperature anisotropies
or the lensing signature by clusters.

In contrast to CMB temperature anisotropies,  CMB polarization anisotropies
have relatively little contamination from the tSZ and kSZ effects;
There are a couple of contamination sources in
CMB polarization anisotropies
such as the scattering of the kinetic quadrupole
in the electron's rest frame, the scattering of the anisotropic CMB photons
from the tSZ effect, and double scattering within the cluster
(e.g., \cite{SUZE80,LEKI06}). However, compared to the lensing effect
$\sim1\muK$ in polarization, 
the effects of these contaminants are negligible $\ll0.1\muK$
as they are proportional to $\beta^2\tau$, $y\tau$, and 
$\beta\tau^2$ with the tangential velocity $\beta=v_t/c\sim10^{-3}$ and 
$y=\tau kT_e/m_ec^2\sim10^{-5}$ (e.g., \cite{AMWH05,SHREET06}). Therefore,
we only consider the kSZ effect as a contaminant in  CMB temperature
anisotropies and no contaminant is assumed in CMB polarization anisotropies
in the remainder of the paper.

Figure~\ref{fig:ksz} illustrates the kSZ effect around the massive cluster
at $z=1$ in Fig.~\ref{fig:stack}. The top panels show the Thomson scattering
optical depth~$\tau$ and the line-of-sight velocity field~$\vlos$ of free
electrons. We assume that the helium mass fraction is $\YHe=0.24$ and 
that gas is fully ionized throughout the simulation box. 
While the gas distribution in massive clusters may have a velocity dispersion
lower than the dark matter distribution, here we simply assume that
the gas distribution traces the dark matter distribution in phase space
with the universal mass fraction $\OB/\Omega_\up{dm}$ and investigate its
impact in Sec.~\ref{ssec:ksz}. Therefore, the top left
panel appears identical to the convergence field (top left)
in Fig.~\ref{fig:stack},
as they are both proportional to the projected mass density. 
The cluster is moving toward the observer at the mean $\vlos=-300~\kms$
and its $\vlos$ distribution is random with the rms $\sigma_v=640~\kms$, 
though the average $\vlos$ seen in Fig.~\ref{fig:ksz}
is smaller due to the inclusion of random interlopers along the line-of-sight
within the finite angular size of detector pixel.

The bottom panels show the kSZ effect $\Delta\Tksz/T=-\tau~\vlos/c$ with
and without the intrinsic CMB temperature field. Most of the intergalactic
medium is transparent ($\tau\simeq0$) and there is little kSZ effect therein,
except a few small blobs along the large-scale filaments. The kSZ effect
$\Delta\Tksz\simeq30\muK$ is highly concentrated at the cluster center
and diminishes fast as the matter density of clusters falls off rapidly.
However, the contamination from the kSZ effect is apparent in the bottom
right panel and its impact increases with redshift as clusters at high
redshift are more compact and the line-of-sight velocity decreases only
with $(1+z)^{-1/2}$.

\section{Reconstructing Cluster-Mass Cross-Correlation}
\label{sec:recon}
Using the numerical simulation in Sec.~\ref{sec:sim}, we demonstrate the
applicability of our improved quadratic estimators to realistic clusters.
In Sec.~\ref{ssec:iqe} we first test five improved quadratic estimators
in an ideal CMB experiment, and we then compare their performance in realistic
CMB experiments. In Sec.~\ref{ssec:ksz}
we discuss the impact of the kSZ contamination on the lensing reconstruction
by using CMB temperature anisotropies and comment on a way to improve the
reconstruction in the presence of the kSZ effect.

\subsection{Improved quadratic estimators using CMB polarization anisotropies}
\label{ssec:iqe}
Here we test our improved quadratic estimators against the numerical
simulation described in Sec.~\ref{sec:sim}.
Figure~\ref{fig:ideal} shows the convergence
profiles reconstructed by applying improved quadratic estimators
(labeled in the legend of each panel) to 500 clusters at $z=1$ in an ideal
experiment with no detector noise ($\spix=0$) and telescope beam 
($\tfwhm=0$). Note that each cluster we stack has different shapes.
We first apply improved quadratic estimators by
adopting an initial convergence model $\Kmod(\theta)=0.02$ (horizontal
dashed), i.e., no lensing signature in the observed CMB temperature and 
polarization anisotropies. Each estimator takes the assumed initial
model and computes the change $\Delta\kappa$ from the assumed model,
yielding an estimate (thin solid)
of the convergence profile.
The first estimates of each of the estimators from the uniform matter distribution 
are already close to the true convergence profile (thick solid).
Having adopted the initial model without lensing effects,
our improved quadratic estimators operate
as standard quadratic estimators, and they are biased low when the lensing
effect is large near the massive clusters \cite{MABAMEET05,HUDEVA07,YOZA08},
while the nonlinear effect is relatively mitigated for iTE, iTB, and iEB
estimators, since the lensing contribution
is partially canceled due to the oscillating nature of the 
cross power spectrum $\CC_l$ and there is no B-mode polarization anisotropy
$\CB_l=0$.

Next we take the estimate (thin line) of the convergence profile in the
first trial as our initial convergence
 model for the next iteration, and we repeat
the iteration process  until the estimates converge.
In just a few iterations, the initial convergence model of the uniform
matter distribution
is quickly reshaped and all the estimators in each panel converge 
to the true convergence profile without any detectable bias beyond the
cluster virial radius (vertical dotted). The uncertainties (shaded region)
in the mean estimate become smaller as the assumed convergence model 
is refined at each iteration;
their minimum is achieved when the estimates
converge to the true convergence profile
and it is set by the intrinsic fluctuations of CMB temperature
and polarization anisotropies. For a faster convergence, one can start with a
more realistic initial model of the convergence field 
motivated by other observations
rather than the uniform matter distribution assumed here. Note that while
we characterize our reconstruction for the cluster-mass cross-correlation
in terms of the convergence profile
$\khat(\theta)$ averaged over the annulus, it is the 2D convergence field
$\khat(\Vang)$ that we reconstruct 
using improved quadratic estimators (similarly for the initial models).

\begin{figure}
\centerline{\psfig{file=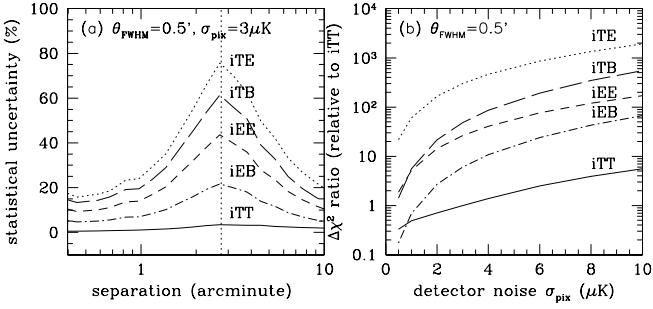, width=3.4in}}
\caption{Relative performance comparison of the improved quadratic 
 estimators for reconstructing
the convergence profile. The left panel shows the statistical uncertainties 
in the mean convergence profile obtained by applying each estimator to
1000 clusters in the fiducial CMB
experiment with $\spix=3\muK$ and $\tfwhm=0.\!'5$.
The right panel shows the
degradation in the total signal-to-noise ratio of each estimator as a function
of the detector noise, compared to that of the iTT estimator 
in the fiducial experiment.}
\label{fig:signoi}
\end{figure}

Figure~\ref{fig:signoi} compares the relative performance of each improved quadratic
estimator in reconstructing the convergence profile in the fiducial
CMB experiment with $\spix=3\muK$ and $\tfwhm=0.\!'5$ and quantifies its
signal-to-noise ratio for different
CMB experiments with varying levels of detector noise.
The left panel plots the statistical uncertainty in the mean estimate of
the convergence profile for each estimator from 1000 clusters,
after a few iterations when the estimates have converged.
The statistical uncertainties are averaged over the logarithmic radial bin,
comparing the {\it relative} performance and they decrease in proportion to
$N_\up{cl}^{1/2}$ with $N_\up{cl}$ being the number of clusters stacked.

The performance difference of each estimator in Fig.~\ref{fig:signoi}a 
originates from their noise power spectrum in Eq.~(\ref{eq:vari}) and 
the noise power spectrum $\Nkap_l$ of each estimator, or equivalently the
normalization $N_l^{XY}$ in Eq.~(\ref{eq:nps}), is determined by the intrinsic
CMB power spectra $C_l^{XY}$, $C_l^X$, and $C_l^Y$, given experimental 
specifications (detector noise and telescope beam size). 
Especially, the relevant information about the lensing reconstruction 
is contained around 
$l\sim1/\theta_\up{cl}\sim3000$. Since the intrinsic CMB power spectra decay
exponentially, the noise power spectrum $N_l^{\kappa,TT}$ for iTT estimators
at $l\gg1/\theta_\up{cl}$
is set by the sum of the ratios $\CT_l/C^{N,T}_l$ at $l\lesssim l_\up{crit}$ and
$l_\up{crit}$ is the scale at which the ratio becomes unity 
($1/\theta_\up{cl}<l_\up{crit}^{TT}$).
Analogously, the relevant ratios can be obtained by
using Eq.~(\ref{eq:nps}) as
$\CE_l/C^{N,E}_l$ for iEB and iEE estimators and $(\CC_l)^2/C^{N,T}_lC^{N,E}_l$ 
for iTB and TE estimators, though the critical scale $l_\up{crit}$ is similar
to the cluster scale, $1/\theta_\up{cl}\simeq l_\up{crit}$ for the polarization
estimators.
Since there is no intrinsic B-mode 
polarization anisotropy, the noise power spectra of iEB and iTB
estimators are somewhat smaller than those of iEE and iTE estimators on cluster
scales. Therefore, the lensing reconstruction of the cluster
convergence profile
can be best achieved by iTT estimators, and iEB, iEE, iTB, and iTE estimators
have larger variance in the sequential order for the fiducial CMB experiment
as shown in Fig.~\ref{fig:signoi}a.

To quantify the signal-to-noise ratio of the improved quadratic estimators
and their dependence on the detector noise, we compute
the covariance matrix of each estimator as
\beeq
\mathcal{C}_{XY}(\theta,\theta')=\left\langle\left[\khat^{XY}(\theta)-
\kappa(\theta)\right]\left[\khat^{XY}(\theta')-\kappa(\theta')\right]
\right\rangle~,
\eneq
and the signal-to-noise ratio for each experiment specifications
is then
\beeq
\Delta\chi^2_{XY}=\sum_{\theta,\theta'}\kappa(\theta)
\left[\mathcal{C}_{XY}(\theta,\theta')\right]^{-1}
\kappa(\theta')~.
\eneq
Since the lensing effect on CMB power spectra 
is to convolve them with its potential power spectrum $\Cphi_l$, the lensing
estimators are intrinsically non-local and the covariance matrix is 
non-diagonal \cite{YOZA08}.
The covariance matrix is computed by averaging over 50,000 clusters to 
guarantee its numerical convergence.

Figure~\ref{fig:signoi}b plots the ratio of $\Delta\chi^2_{TT}$ 
of the TT estimator
in the fiducial experiment to $\Delta\chi_{XY}^2$ 
of each estimator as a function
of detector noise. As in Fig.~\ref{fig:signoi}a,
the general trend of the performance of each estimator
remains unchanged over a wide range of detector noise; as the detector noise 
increases, the performance of all the estimators is progressively degraded.
However, the signal-to-noise ratios dramatically improve at $\spix<3\muK$
for the estimators using CMB polarization anisotropies, since at the
cluster scale the intrinsic CMB polarization anisotropies are small and
easily dominated by the detector noise.

\begin{figure}
\centerline{\psfig{file=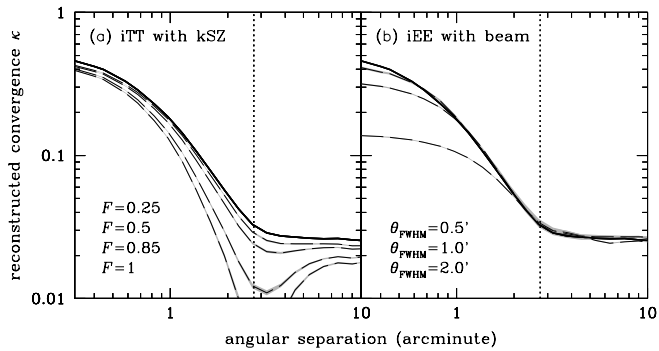, width=3.4in}}
\caption{Impact of the kinetic Sunyaev-Zel'dovich (kSZ) effect on the lensing
reconstruction. For the fiducial experiment with $\tfwhm=0.\!'5$ and
$\spix=3~\muK$, we use the simulation in Sec.~\ref{sec:sim} to generate
10,000 massive clusters at $z=1$ and stack the 
convergence profiles reconstructed by
applying iTT and iEE estimators. Left panel:
Assuming that the gas distribution traces the dark matter 
distribution in phase space from the simulation,
we compute the kSZ contamination as a function of the ionized gas fraction~$F$
in Eq.~(\ref{eq:frac}). 
The reconstructed convergence profile is progressively biased
in the presence of the kSZ contamination in CMB temperature anisotropies.
Right panel: Convergence profiles with iEE estimators as a function
of telescope beam. No contamination from the clusters is assumed in CMB
polarization anisotropies. Reconstructed convergence profiles are shown
as dashed lines with their uncertainty (shaded), while the true convergence
profile is shown as solid lines.}
\label{fig:kszimpact}
\end{figure}

\subsection{Kinetic Sunyaev-Zel'dovich effect}
\label{ssec:ksz}
Now we use the simulation in Sec.~\ref{ssec:sec} with its velocity distribution
to investigate the effect of the kSZ contamination illustrated 
in Fig.~\ref{fig:ksz}
on the lensing reconstruction. Note that we
consider only the kSZ effect as a possible contamination of CMB temperature
anisotropies and no other contamination is assumed for CMB polarization 
anisotropies, except the lensing effect itself (see, Sec.~\ref{ssec:sec}
for estimates of possible contamination in CMB polarization anisotropies).

In massive clusters, baryons make up a universal mass fraction $\OB/\OM=0.138$ 
and they exist predominantly in the form of ionized gas in the intracluster
medium. However, stars and galaxies in massive clusters contain a non-negligible
fraction of baryons, and X-ray observations show that the gas mass fraction is
$f_\up{gas}=0.117$ within $r_{2500}$, which corresponds to $\sim0.1\Rvir$:
Approximately 85\% of the baryons in massive clusters contributes to the kSZ
effect (see, e.g., \cite{ALSCFA02,VIKRET06} for estimating the 
gas mass fraction).
Here we simply adopt the ionized gas fraction~$F$ as a free parameter
and the free electron number density is then obtained from the simulation by
\beeq
n_e=F\times\left({\OB\over\Omega_\up{dm}}\right)
\left[{m_\up{dm}\over m_p}\left(1-{\YHe\over2}\right)\right]~,
\label{eq:frac}
\eneq
where $m_p$ is the proton mass. Since the kSZ effect is proportional to
the product of the free electron number density and the line-of-sight
velocity, a lower value of~$F$ can represent a lower velocity distribution
of baryons in massive clusters with the universal mass fraction 
$\OB/\OM$.

Figure~\ref{fig:kszimpact}a shows the impact of the kSZ contamination
on the lensing reconstruction (dashed) from iTT estimators in the fiducial 
experiment. When most baryons are contained in the ionized gas
($F=1$ and $F=0.85$), the kSZ contamination is $\Delta\Tksz\simeq30\muK$ 
and~$25\muK$ at the cluster center, respectively, 
and its effect is substantially
larger than the typical change $\Delta T\simeq10\muK$ arising from the lensing 
effect by clusters. Consequently, the lensing reconstruction is significantly
biased in the presence of the kSZ contamination. For the other two 
cases, in which large fraction of the baryons are in stars ($F=0.5$ and $F=0.25$),
the kSZ contamination $\Delta\Tksz=15\muK$ and~$7.5\muK$ is smaller and its
impact is reduced, though the bias in the lensing reconstruction
prevails over a range of separations.

Compared to CMB temperature anisotropies, CMB polarization anisotropies are 
relatively free from secondary contamination and the lensing effect is the
dominant source of secondary anisotropies. However, the current CMB experiments
such as SPT and ACT lack the ability to measure polarization anisotropies,
and other 
upcoming CMB experiments may not be optimized for measuring the cluster
lensing signature in polarization anisotropies.
Figure~\ref{fig:kszimpact}b illustrates the lensing reconstruction by using
iEE estimators in future CMB polarization experiments with larger telescope 
beam size (B-mode polarization measurements are even harder due to the
vanishing signal).
No significant bias develops for iEE estimators within the virial
radius. However, as the angular resolution decreases, small scale structure 
is smoothed and the reconstructed convergence profile is the 
true convergence profile convolved with the telescope beam (dashed line).

Finally, with the current CMB experiments capable of measuring only temperature 
anisotropies on cluster scales, we consider a way to mitigate the impact
of the kSZ contamination in the lensing reconstruction using CMB temperature
anisotropies. Since the kSZ effect arises from the Thomson scattering off free 
electrons in clusters, its contamination is centrally concentrated as seen
in Fig.~\ref{fig:ksz}, scaling in proportion to $\kappa(\theta)$. The lensing
effect, however, extends well beyond the innermost region of clusters; for 
example, the kSZ effect falls off as $1/\theta$ in singular isothermal
clusters, while the lensing effect remains constant throughout the cluster
region. Therefore, the simplest way to reduce the kSZ contamination is to suppress
or purge the central region of clusters, where
the kSZ effect is strongest (see, e.g., \cite{VAAMWH04}).
Since the lensing estimators are non-local and
the information on the central region is shared by lensed CMB temperature 
anisotropies over the cluster virial radius \cite{YOZA08},
the lensing reconstruction is possible even with the pixels at the innermost
region of clusters masked out. However, with our quadratic estimators built in
Fourier space, translational invariance prevents from discriminating the central region.
We leave the further investigation of how masking the central region 
affects the performance of our estimator for future work.

\section{Discussion}
\label{sec:dis}
We have generalized the maximum likelihood estimators \cite{YOZA08}
for reconstructing 
the cluster-mass cross-correlation
 to utilize delensed CMB polarization fields, and we 
have tested our new maximum likelihood estimators for the lensing 
reconstruction against numerical simulations. The observed CMB temperature
and polarization fields are delensed based on an initial model of the 
convergence field and our maximum likelihood estimators (called improved
quadratic estimators) provide improvements over the assumed initial model. 
This process can be iterated until numerical convergence is achieved.
Compared to the standard and modified quadratic estimators
\cite{MABAMEET05,HUDEVA07}, our improved quadratic estimators have no free
parameters and provide unbiased reconstructions even in the regime 
where the linear approximation in the lensing effect breaks down.

We have adopted smoothed hydrodynamic simulations
to model realistic clusters,
and the improved quadratic estimators that use six different combinations
of CMB temperature and polarization fields can reconstruct the underlying matter
distribution in a non-parametric way to obtain the cluster-mass 
cross-correlation. Their ability to reconstruct the
cluster-mass cross-correlation is determined by the ratio of the intrinsic
CMB power spectrum to the detector noise power spectrum on cluster scales.
Given the same experimental specifications for measuring CMB temperature and
polarization anisotropies, iTT estimators can reconstruct the cluster-mass
cross-correlation by a factor of~3 in the signal-to-noise ratio, better
than iEB estimators, which are better suited for reconstructing large-scale
structure.

For the gas mass fraction $f_\up{gas}=0.117$, the numerical simulations show
that the kinetic Sunyaev-Zel'dovich (kSZ) contamination is 
$\Delta\Tksz\simeq30\muK$
at the innermost region of clusters. As clusters are more compact at high
redshift with weak redshift dependence of their peculiar velocity, the 
spectrally indistinguishable kSZ contamination in CMB temperature anisotropies 
poses a significant challenge to the lensing reconstruction 
using iTT estimators. However,
polarization estimators such as iEB estimators
or more practically iEE estimators can be used for the 
cross-check of the lensing
reconstruction. Especially, these polarization estimators are more desirable
given that there is relatively little kSZ contamination in CMB polarization
anisotropies and they perform better than iTT estimators in an ideal CMB
experiment with no detector noise and telescope beam.
However, with the prospect of CMB polarization experiments competitive 
with the current CMB temperature experiments far in the future, 
masking the central region of clusters to mitigate the kSZ contamination
will be needed to take full advantage of precise CMB temperature measurements
in the lensing reconstruction. 

\acknowledgments 
J.~Y. acknowledges useful discussions with Adam Lidz, Jonathan Pritchard,
and Amit Yadav.
J.~Y. is supported by the Harvard College 
Observatory under the Donald~H. Menzel fund.
M.~Z. is supported by the David and Lucile Packard, the Alfred~P. Sloan,
and the John~D. and Catherine~T. MacArthur Foundations.
This work was further supported by NSF grant AST~05-06556 and NASA ATP
grant NNG~05GJ40G.

\vfill

\bibliographystyle{/home/jyoo/Journal/Latex-Style/apsrev}
\bibliography{/home/jyoo/Journal/Manuscripts/reference}

\end{document}